\documentstyle[epsf,aps,multicol]{revtex}

\begin{document}

\title{External-field-induced tricritical point in a
fluctuation-driven nematic--smectic-A
transition}

\author{Ranjan Mukhopadhyay, Anand Yethiraj, and John Bechhoefer}
\address{Department of Physics, Simon Fraser University, Burnaby, British
Columbia, V5A 1S6, Canada}


\maketitle

\begin{abstract}
We study theoretically the effect of an external field on the 
nematic--smectic-A (NA) transition close to the tricritical point, 
where fluctuation effects govern the qualitative behavior of the 
transition.  An external field suppresses nematic director 
fluctuations, by making them massive.  For a fluctuation-driven 
first-order transition, we show that an external field can drive the 
transition second-order.  In an appropriate liquid crystal system, we 
predict the required magnetic field to be of order $10 T$.  The 
equivalent electric field is of order $1 \, V / \mu {\rm m} $.
\end{abstract}


\begin{multicols}{2}
One of the seminal developments in the theory of phase transitions was 
the understanding of how thermal fluctuations can change the apparent 
analytic properties of the free energy and thereby render the 
predictions of mean-field theories invalid.  The most widely 
appreciated consequence of thermal fluctuations is the shift in 
critical exponents from their mean field values \cite{goldenfeld}.  Another kind of 
consequence, less widely known, was predicted over two decades ago by 
Halperin, Lubensky and Ma \cite{HLM}.  They argued that the coupling between the 
fluctuations of a gauge field with the order parameter can convert a 
second-order transition to a first-order one (the ``HLM effect'').  
Such fluctuation-induced first-order transitions are expected in two 
systems: the BCS transition in type-1 superconductors and the 
nematic--smectic-A (NA) transition of liquid crystals.  In 
superconductors, the HLM effect is immeasurably weak; however, in liquid 
crystals, Anisimov, Cladis, and coworkers \cite{anisimov2,cladis1} have 
found experimental evidence that is consistent with it.  But as we 
discuss below, that evidence is indirect.  In this paper, we show 
that applying a modest 
external field along the preferred orientation of the nematic leads to 
effects that may be unambiguously attributed to the coupling proposed 
by HLM, providing a direct test of the HLM scenario.

In liquid crystals, the HLM effect is sensitive to the ratio 
$T_{NA}/ T_{NI}$, $T_{NA}$ and $T_{NI}$ being the nematic--smectic-A 
(NA) and nematic-isotropic (NI) transition temperatures, respectively.  
When the nematic range is large, i.e., when the NI transition is 
sufficiently far from the NA transition ($T_{NA}/ T_{NI}\ll1$), the 
transition is expected to be second order.  Indeed, experiments on 
such systems yield critical exponents consistent with the 3D XY model 
\cite{dejeu2,garland3} and show no detectable discontinuities.


For small nematic range ($T_{NA}/T_{NI} \rightarrow 1$), the nematic 
order parameter, which increases sharply on cooling below $T_{NI}$, 
has not yet saturated when the NA transition is reached.  The nematic 
phase is thus only partially ordered at the NA transition, and the 
emerging smectic order parameter $\psi$ is intrinsically coupled to 
both the nematic order parameter magnitude $S$ ($\delta S - \psi$ 
coupling) and the fluctuations in the director $\hat{\bf{n}}$ ($\delta 
{\bf{n}} - \psi$ coupling).  
%

There is some evidence for a crossover from a first-order transition 
driven by $\delta S - \psi$ coupling to one driven by $\delta {\bf n} 
- \psi$ coupling, as predicted by the HLM theory.  Experimentally, 
$T_{NA}/T_{NI}$ can be tuned by mixing liquid crystals with slightly 
different aliphatic chain lengths.  Based on the $\delta S - \psi$ 
coupling, one expects that in a mixture with mole-fraction $x$, the 
latent heat $L$ at the NA transition goes as $L \propto x - 
x^{\star}$, where $x^{\star}$ is the mole-fraction for a mixture at 
the Landau tricritical point (LTP).  In one such mixture (of two 
cyanobiphenyls, 8CB and 10CB), Marynissen and coworkers \cite{maryn1} 
have found such a linear dependence.  The extrapolation of this linear 
behavior gives the position of the LTP to be at $40\%$ mole-fraction 
of 10CB in 8CB.  However, the latent heat does not go to $0$ at the 
LTP.  Instead, there is a quadratic crossover to a smaller 
value for $x < x^{\star}$.  Anisimov and coworkers interpreted this 
result\cite{anisimov2} as evidence supporting the HLM theory \cite{HLM}.  
Independent work 
measuring the capillary length of 8CB-10CB mixtures by Tamblyn {\it et 
al.}\cite{nancy1} shows a similar crossover.  Moreover, even for pure 
8CB ($x = 0$), Cladis {\it et al.} \cite{cladis1} have deduced from 
front propagation experiments that the transition is first order.  
This has been confirmed by Yethiraj and Bechhoefer 
\cite{yethiraj}, who measured nematic fluctuations directly in real space.

Although these experiments suggest the existence of the non-analytic 
cubic term in the smectic free energy, they do not show unambiguously that 
this effect arises from the HLM mechanism. 
One can directly probe the effect of director fluctuations on the 
nature of the transition by expanding the parameter space of the 
free energy to include an external magnetic (or electric) field. 
As we shall see below, the HLM theory, thus modified, gives rise to a peculiar 
form for the external-field dependence of measured quantities.  An 
experimental observation of this specific form would be hard to 
attribute to any other mechanism.

In addition, applying an external field 
affords an experimentalist two other opportunities:  first, direct suppression 
of fluctuation effects provides a continuously variable parameter with 
which to study the approach to the tricritical point in a single 
material.  In contrast, each data point in mixtures corresponds to a 
different concentration and is therefore a different experiment.  More 
important, mixtures may differ in properties other than simply the 
ratio of ${T_{NA}/T_{NI}}$, which complicates the comparison of 
different experiments.  Second, the external field provides a way of 
suppressing 
the anisotropic coupling that gives the correlation-length exponents 
at the NA transition their weak anisotropy.  In what follows, we will 
show that the subtle fluctuation effects at play at the NA transition 
can be tuned by modest magnetic (or electric) fields, making 
concrete predictions that can be checked experimentally.

We start with the free energy proposed by de Gennes. Because the 
nematic phase from which the smectic condenses is only partially ordered,
the free-energy expansion must consider the effects of both 
nematic and smectic ordering.
As the density modulation in smectic
liquid crystals is nearly harmonic, one may write
$\rho({\bf{r}}) \simeq {\rho_0}({\bf{r}})[1 + \rho_{1}({\bf{r}})
   \sin(\bf{q}_0 \cdot
 \bf{r} - \phi)]$ and  define the smectic order parameter by
$ \Psi({\bf{r}}) \equiv \rho_{1}({\bf{r}})e^{i\phi({\bf{r}})}$.
Assuming that $\hat{\bf n}$ fluctuates about the $z$-axis,
one can write the free energy as \cite{dGP}
\begin{eqnarray}
F_{NA} & =& \int d^3 x f_{NA}(\Psi, \delta {\bf n})=
{1 \over 2} \int d^{3} x \{\bar{r}|\Psi|^2 + {u \over 2}|\Psi|^4 \nonumber \\
& &+C_{\parallel}\left|{\frac{\partial\Psi}{\partial z}}\right|^2 
+ C_{\perp}\left|(\nabla_{\perp} - i q_0  \delta{\bf n}_{\bot}
)\Psi\right|^2 \nonumber \\
& & + K_1 (\nabla \cdot\delta {\bf{n_{\perp}}})^{2} + K_2 ( \hat{z} \cdot
\nabla \times \delta {\bf{n}}_{\bot})^2  \nonumber \\
& &+ {K_3({{\partial} \over
{\partial {z}}} \delta {\bf{n}}_{\bot})^2} \},
\end{eqnarray}
where $\delta {\bf n}_{\perp} = ( \delta n_{x}, \ \delta n_{y}, \ 0)$.
We rescale lengths in the $\hat{z}$ direction 
relative to other directions so that $C_{\perp}=C_{\parallel}=C$
and thus the three Frank constants $K_{1,2,3}$ are also rescaled. 
Close to the NA transition,
$\bar{r}$ has the usual form $\alpha{(T-T_0) \over T_0}$. 
The last three terms in the free energy correspond to the
splay, twist, and bend contributions to the Frank elastic
energy for nematics.  
Note that we have not explicitly included the coupling 
between the smectic order parameter $\Psi$
and nematic order parameter $S$, as its main effect 
 is to shift $\bar{r}$ and $u$. We thus use effective
values of $\bar{r}$ and $u$.
In the absence of $\delta \bf{n}$ fluctuations,
$u = 0$ corresponds to the tricritical point, and $u > 0$ implies a 
second-order transition. 
However, when $\delta\bf{n}$ fluctuations are taken into account, 
nothing special happens at $u=0$: we merely cross-over from a 
mean-field first-order transition to a fluctuation-driven first-order 
transition.
Here, we assume $u \geq 0$.

Next, we consider the effect of an external field
along the director $\hat{\bf{n}}$ (assumed to lie along the z-axis).
 We assume that the field
reinforces the nematic ordering and neglect its much
smaller effects on smectic ordering.
For concreteness, we consider a magnetic field $H$.
Then
the Landau free energy becomes
\begin{eqnarray}
F^{H}_{NA} &=& F_{NA} - \int d^3 x \,{1 \over 2}\chi_a 
({\bf{H}} \cdot \hat{\bf{n}})^2 \nonumber \\
&\simeq& F_{NA} - {1 \over 2}V \chi_a H^{2} 
       + \int d^{3}x {1 \over 2} \chi_a H^2 \delta n^{2} ,
\end{eqnarray}
using ${n_{z}}^2=(1 - \delta{\bf n}^2)$, and expanding in $\delta n^{2}$.
Here, $V$ is the sample volume.
Thus, the magnetic field makes the nematic
director fluctuations ``massive''. Because the field also couples to
the nematic order parameter $S$, 
the free-energy-expansion  coefficients also
have a magnetic-field dependence \cite{rosenblatt1,hama}; however, one would
need a field of several hundred Teslas
 to change $u$ appreciably, which  is a much weaker effect
than the one we shall be considering here.

Recall that we are working
in the regime where massless nematic director fluctuations,   
by coupling to the smectic order parameter, induce a 
first-order transition. Adding an external field
adds mass to these director fluctuations, thus suppressing
their effect. When the magnetic field is sufficiently strong,
director fluctuations can be ignored,  resulting in a 3D-XY,
second-order transition. 
To estimate the required critical field, we recall that nematic 
twist and bend distortions 
are expelled by the smectic phase over a length scale
$\lambda$ (defined to be the penetration depth). At a mean-field level,
and in the one-constant approximation $ K_1 = K_2 = K_3 = K$,
the penetration length is given by
\begin{equation}
\lambda = \left({K \over C}\right)^{1/2}{1 \over {q_0 |\Psi_0|}}.
 \label{eq:l1}
\end{equation}
When a field is added,
we introduce a new length, the magnetic coherence length $\xi(H)$,
which  measures  the distance over which elastic deformations
 decay  in the nematic phase. One finds \cite{dGP}
\begin{equation}
\xi(H) = \left({{K} \over {\chi_a}}\right)^{1 \over 2}
{1 \over H}. \label{eq:l2}
\end{equation}
At zero field, if the transition is first order, we can imagine 
smectic droplets in the nematic phase
at the coexistence temperature.  Bulk twist and bend excitations
penetrate a distance $\lambda$ into the smectic droplets. 
When $H$ is turned on, as long
as $\xi(H)$ is much larger than $\lambda$, 
the nematic-smectic interface is not much affected.
But when $\xi(H)$ is much smaller
than $\lambda$, nematic fluctuations are  suppressed in
both the nematic and smectic phases. They then play no role
at the transition, which becomes second-order XY.
Thus, a rough estimate of magnetic field $H_c$ needed to reach the
tricritical point can be obtained by setting $\xi(H_c) = \lambda$.
 In reality,
the different values of $K_1$,
$K_2$, and $K_3$  lead to
different penetration depths and magnetic
coherence lengths for the twist and bend modes, somewhat
complicating the above arguments. 

We can study the effect of a magnetic field 
within the Halperin, Lubensky and Ma (HLM) formalism, where  
fluctuations in $|\Psi|$ are ignored (the strongly ``type-1'' limit). 
In order to decouple fluctuations in the phase of the order parameter, 
$\Psi$,
from director fluctuations, we carry out the gauge transformation
\cite {HLM2},
$\delta {\bf n}_{\perp} = {\bf A} + \nabla L$, and 
$\psi = \Psi e^{-i {q_0} L}$, where $\nabla \cdot {\bf A}= 0$.
Under this transformation, $|(\nabla - i {q_0} \delta {\bf  
n}_{\perp}) \Psi|^{2}$ goes to $(\nabla - i {q_0} {\bf A})
\psi|^2$. Details  will be   
given in a longer paper \cite{mukho}. Following HLM, we write
\begin{equation}
e^{- F(\psi)/k_B T} = \int {\mathcal{D}} \{ {\bf A} \} 
   e^{- {F_{NA}(\psi, {\bf A})}\over {k_B T}} \ .
\end{equation}
Differentiating with respect to $|\psi|$ gives
\begin{equation}
 {{df} \over{d|\psi|}} = {\bar{r}}|\psi| + u|{\psi}|^3 + C q_{0}^2 |\psi|
\langle {\bf A}^{2} \rangle .
\end{equation}
For simplicity, we will assume that we are in the limit $K_1 \ll K_2, K_3$, and
hence set $K_1$ to zero \cite{note1}, however the results would not change
much for finite $K_1$. 
Treating $|\psi|$ as a constant, we obtain
\begin{eqnarray}
{{df \over d|\psi|}} &=& r|\psi| + u|\psi|^3 - {w_1}|\psi|{\sqrt{(
|\psi|^2 + a_H^{2} H^2)}} \nonumber \\
 &+& {w_2}{a_H}|\psi||H|\ln\left[ {{\sqrt{|\psi|^{2}
+ a_{H}^{2}H^{2}} + {a_H}|H|} \over {2 {a_H} |H|}} \right] \ ,
\end{eqnarray}
where
$w_{1} = {{k_B T}\over{\pi}} {{C^{3/2} q_0^{3}} \over {2 
K_3^{1/2}}} \left({1 \over K_3} + {1 \over K_2}\right)$, 
$w_{2}= {{k_B T}\over{\pi}} {{C^{3/2} q_0^{3}} \over {2 
K_3^{3/2}}}$,
$a_H = 
\sqrt{{\chi_a}\over{C_{\perp}q_0^{2}}}$, and $r$ corresponds to a 
shift of $\bar{r}$.  
At this point, it is convenient to introduce the scaled (dimensionless)
variables $|\psi^{'}| = {u \over w}|\psi|$, 
${r}^{'}  = {u \over {w^2}}r$,
$H^{'}  = {{u a_{H}} \over w}H$, 
$f_{NA}^{'} = {{u^3} \over {w^4}} f_{NA}$. In terms of these 
variables, the scaled effective free energy density takes the form
\begin{eqnarray}
f^{'} &=& ({r^{'} \over 2} - {b \over 4}|H^{'}|)
|\psi^{'}|^{2} + {1 \over 4}|\psi^{'}|^{4}
 - {1 \over 3}\sqrt{(|\psi^{'}|^{2} + {H^{'2}})^{3}} \nonumber \\
      & & + {b \over 2} H^{'2}\sqrt{(|\psi^{'}|^{2} + {H^{'2}})} \nonumber \\
      & &  + {b \over 2} |H^{'}||\psi^{'}|^{2} \ln\left[ {{\sqrt{|\psi^{'}|^{2}
 + H^{'2}} +  |H^{'}|} \over {2  |H^{'}|}} \right],
\end{eqnarray}
where $b = {{K_2} \over {K_2 + K_3}}$, and $0 < b < 1$. 
Qualitatively, one sees that as $H \rightarrow 0$ there is a 
negative $|\psi|^3$ term, indicating a first-order transition;
for large $|H|$ the last three terms in Eq. 8 give only corrections
to $|\psi|^2$ and $|\psi|^4$, implying a second-order transition.
Thus, we expect a tricritical point at $H_c \simeq {\psi}_{0}/a_H
\simeq {\frac{1}{a_H}}{w \over u}$.

At the coexistence point, using $f(|\psi_0|) = f(0)$ and
$\left[{{df} \over {d|\psi|}}\right]_{\psi=\psi_{0}}=0$, we obtain 
\begin{eqnarray}
| \psi_0^{'}|^{2} &= {{(2-9b|H^{'}|) +  {\sqrt{\Delta}}}\over 9}
     & \ \ {\mathrm{for}} \  0<H^{'}<{1 \over 3}  \nonumber \\
     &=  {{(2-9b|H^{'}|) -  {\sqrt{\Delta}}}\over 9} 
     & \ \ {\mathrm{for}} \  {1 \over 3} < H^{'} <{H^{'}_c},
\end{eqnarray} 
where 
\begin{eqnarray}
\Delta&=&(9b |H^{'}| -2)^{2} \nonumber \\
      & &  - 108(1-3b/2){H^{'2}}[(1 - b/2) - 2 |H^{'}|] .
\end{eqnarray}
Here $H^{'}_c={1 \over 2}(1 - b/2)$ is the critical field; at
$H = H_c$ we have $\psi_{0} = 0$ at the coexistence point. Thus, the
tricritical point corresponds to $H=H_{c}$, for larger magnetic
fields the NA transition is second-order.
Our earlier informal argument giving $\lambda_{H=0}=
\xi_{H_c}$ 
corresponds to $H_c^{'} = {2 \over 3}$. Note that,
despite appearances, $\psi_{0}(H)$ is analytic at $H^{'}= 1/3$.
(See Fig.~1, inset.)
The coexistence temperature $r_{NA}$ satisfies
\begin{eqnarray}
r_{NA}^{'} &=& - |\psi^{'}_0|^{2} + {\sqrt{|\psi^{'}_0|^{2} + 
          {H^{'2}}}} \nonumber\\
            & &   - b|H^{'}|\ln\left[ {{\sqrt{|\psi^{'}_{0}|^{2} 
+  H^{'2}} + |H^{'}|} \over {2  |H^{'}|}} \right] .
\end{eqnarray}

Because the spinodal temperature $T^{*}$ changes linearly in
$|H|$,  there is  a cusp at $H=0$ in the function $t_{0}(H)$.
(See Fig. 1.) The  behavior of $t_0(H)$ near zero field
is the non-analytic ``signature'' of the field-driven HLM 
effect in the same way that a $|\psi|^3$ term is the signature of the 
zero-field HLM effect. Recall that
only $H^{2}$ figures in the original free energy. It should also
be noted that $t_0$ as a function of $H/H_{c}$ does not depend
significantly on $b$. 
\begin{figure}
\par\columnwidth20.5pc
\hsize\columnwidth\global\linewidth\columnwidth
\displaywidth\columnwidth
\epsfxsize=3.0truein
\centerline{\epsfbox{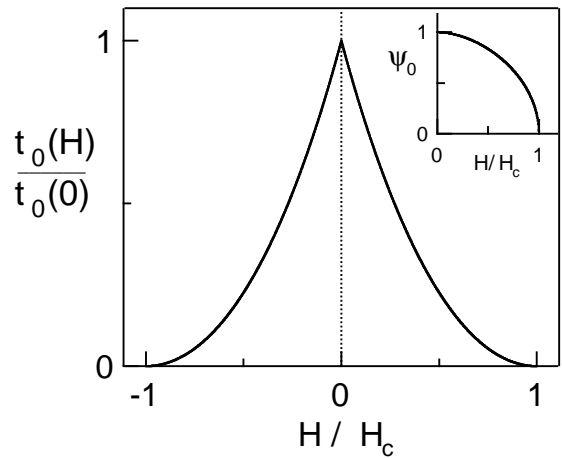}}
\caption{ Plot of the reduced temperature $t_0(H)={{T_{NA} - T^{*}} \over 
T_{NA}}$ as
a function of the scaled magnetic field $H$,
at $b=0$. Note the cusp at $H=0$. In 
the insert, we plot the (scaled) smectic order parameter at the transition 
as a function of $H$.}
\label{fig:EF2}
\end{figure}

To estimate the magnitude of magnetic field required to drive the 
transition second order,  we consider the material 8CB, where the NA 
transition appears to be in the HLM 
fluctuation-driven first-order regime.  
It is useful to express $H_c$
in terms of the measured value of $t_0$ at zero field.
In the HLM formalism, $t_0 = {2 w^{2}
\over 9 \alpha u}$, and we have
\begin{equation}
H_c = \left[{9 \over 8}{\alpha \over u} {C_{\perp} q_{0}^{2}
\over \chi_a}\right]^{1/2}(1 - b/2) {\sqrt{t_0}} \equiv H_0 \sqrt{t_0} \ .
\end{equation}
Using $C_{\perp} = 2 \times 10^{-7}$ dynes, $q_{0} = 2 
\times 10^{7} {{\mathrm cm}^{-1}}$, $\chi_{a} = 10^{-7}$ c.g.s.,
 $\alpha / u = 1$, we 
estimate $H_0 \approx 3500(1 - b/2)$ Teslas, which is the field 
required to quench fluctuations at molecular scales.
 Using $t_0 = 6 \times 10^{-6}$ \cite{yethiraj}, we obtain the critical field
$H_c \approx$ 5-10 Teslas. For an electric field, the critical electric
field is roughly 0.5-1 V/$\mu {\rm m}$.     

These figures are encouragingly low, but one should be cautious
since smectic fluctuations, which we have 
ignored, are important for such weak first-order transitions.  
The calculation
of the critical magnetic field is on firmer ground in the vicinity of 
the Landau tricritical point ($u = 0$), where the neglect of 
$\psi$ fluctuations is more valid. 
Close to the tricritical point,
we retain a $v {{|\psi|^6} \over 6}$ term in the Hamiltonian.
Then the critical field for a second-order transition is
$ {1 \over a_H} ({w \over v})^{1/3} $.  The expression for $H_{c}$ in 
this case is
\begin{equation}
H_c \approx \left[\left({ \alpha \over v}\right)^{1/2}{C_{\perp} q_{0}^{2}
\over \chi_a}\right]^{1/2} \;{{t_0}^{1/4}} 
\end{equation}
In the 8CB-10CB system studied by several 
groups\cite{maryn1,nancy1}, the LTP
occurs at a mole fraction of roughly $40 \%$ 10CB in 8CB.  In this system,
one of us has measured $t_{0}$ to be roughly 
$10^{-4}$ \cite{yethirajthesis}.
Unfortunately, the ${t_{0}}^{{1/4}}$ dependence then results in a much
higher critical field, on the order of 300 T ( or roughly 30 V/$\mu {\rm 
m}$). 

Lelidis and Durand have extensively studied 
the effects of large
electric fields on the NA transition \cite{lelidis1,lelidis4}.
In his Ph.D. thesis, Lelidis
looked for evidence of an electric field-induced tricritical point
at the NA transition of 8CB. The experiments, which measure $S$,
give some evidence for a tricritical point at an external electric
field somewhere between 5 and 20
V/$\mu {\rm m}$. Unfortunately the temperature resolution was 25 mK. Since
the zero-field discontinuity is only 2 mK, this does not rule out a 
much smaller critical field for 8CB, and better temperature
resolution will be needed to confirm these results.

External fields may have other interesting,
observable effects.
In the type-2 limit, where the NA transition should be second
order, applying a field
should be  a relevant perturbation that
changes the universality class of the transition
to second-order XY. In the
superconductor analogy, adding a field in the liquid crystal
system corresponds to adding mass to the gauge fluctuations
in superconductors. For massless fluctuations, in type-2
superconductors, magnetic vortices are screened by current
loops. The 
interaction between the current loops (and not the vortices) 
is long range, giving rise to the inverted XY transition. However,
when the gauge fluctuations become massive, the interaction between
current loops decays exponentially while that between
vortices becomes long range, leading to the usual XY transition.
It would be very interesting to probe the
experimental consequences of this crossover to the XY
fixed point. 

One such consequence
would be the suppression of spatial anisotropy
in the critical region. It has been proposed that 
experiments probe the crossover region between 
an isotropic, high-temperature region and the 
true critical region governed by a
renormalization group fixed point \cite{patton1}.
The nature of the fixed point is still under debate.
 Since with a magnetic
field we could tune the strength of nematic fluctuations, 
it would give us a better understanding of the role of
these fluctuations in the crossover region, and the
observed weak anisotropy. This is currently under investigation.  

In conclusion, we have shown that the HLM effect for the NA 
transition leads to an unusual, non-analytic form for the effective 
smectic free energy in the presence of an external field.  
Measurements by Lelidis and Durand seem consistent with the predicted 
effects.  More precise experiments on the field dependence would be an 
extremely promising way to probe these unusual effects of thermal 
fluctuations.

This work was supported by NSERC (Canada).  
We acknowledge useful discussions with Ian Affleck and Jacques Prost. 


%
%
%


\begin{thebibliography}{10}

\bibitem{goldenfeld}
N. Goldenfeld, {\em Lectures on Phase Transitions and the Renormalization 
Group}, 1st  ed. (Addison-Wesley, Reading, 1992).

\bibitem{HLM}
B.~I. Halperin, T.~C. Lubensky, and S.~K. Ma, Phys.\ Rev.\ Lett.\ {\bf 32},
  292  (1974).

\bibitem{anisimov2}
M.~A. Anisimov {\it et~al.}, Phys.\ Rev.\ A {\bf 41},  6749  (1990).

\bibitem{cladis1}
P.~E. Cladis {\it et~al.}, Phys.\ Rev.\ Lett.\ {\bf 62},  1764  (1989).

\bibitem{dejeu2}
W.~G. Bouwman and W.~H. de~Jeu, Phys.\ Rev.\ Lett.\ {\bf 68},  800  (1992).

\bibitem{garland3}
C.~W. Garland, G. Nounesis, and K.~J. Stine, Phys.\ Rev.\ A {\bf 39},  4919
  (1989).

\bibitem{maryn1}
H. Marynissen, J. Thoen, and W. van Dael, Mol. Cryst. Liq. Cryst. {\bf 124},
  195  (1985).

\bibitem{nancy1}
N. Tamblyn, P. Oswald, A. Miele, and J. Bechhoefer, Phys.\ Rev.\ E {\bf 51},
  2223  (1995).

\bibitem{yethiraj}
A. Yethiraj and J. Bechhoefer, Mol. Cryst. Liq. Cryst. {\bf 304},  301  (1997).

\bibitem{bartholomew}
J. Bartholomew,  Phys.\ Rev.\ B {\bf 28,}  5378  (1983).

\bibitem{dGP}
P.~G. de~Gennes and J. Prost, {\em The Physics of Liquid Crystals}, 2nd  ed.
  (Clarendon Press, Oxford, 1993).

\bibitem{rosenblatt1}
C. Rosenblatt, J. Phys. (Paris) Lett. {\bf 42},  L9  (1981).

\bibitem{hama}
H. Hama, J. Phys. Soc. Jpn. {\bf 54},  2204  (1985).

\bibitem{HLM2}
B.~I. Halperin and T.~C. Lubensky, Solid State Commun.
 {\bf 14}, 997  (1974).

\bibitem{mukho}
R. Mukhopadhyay, (unpublished).

\bibitem{note1} 
The limit $K_1 \rightarrow 0$ should be taken carefully.
The value of $\langle {\bf A}^{2} \rangle$ diverges in
this limit, however the divergent piece corresponds to
a shift of $r$ and does not affect the calculation. 
 

\bibitem{yethirajthesis}
A. Yethiraj, Ph.D. thesis, Simon Fraser University, 1999.

\bibitem{lelidis1}
I. Lelidis, Ph.D. thesis, Universit\'e de Paris-Sud U.F.R. Scientifique
  D'Orsay, 1994.

\bibitem{lelidis4}
I. Lelidis and G. Durand, Phys.\ Rev.\ Lett.\ {\bf 73},  672  (1994).


\bibitem{patton1}
See, for example,
B.~R. Patton and B.~S. Andereck, Phys.\ Rev.\ Lett.\ {\bf 69},  1556  (1992);
B.~S. Andereck and B.~R. Patton, Phys.\ Rev.\ E.\ {\bf 49},  1393  
(1994).

%


\end{thebibliography}

\end{multicols}
\end{document}